# SOLAR NEUTRINOS (almost) WITHOUT STANDARD SOLAR MODELS

G. Fiorentini$^a$ and B. Ricci$^b$ *

$^a$ *Dipartimento di Fisica, Università di Ferrara, I-44100 Ferrara, Italy
and Istituto Nazionale di Fisica Nucleare, Sezione di Ferrara, I-44100
Ferrara, Italy.*
$^b$ *Scuola di Dottorato, Università di Padova, I-35100 Padova, Italy
and Istituto Nazionale di Fisica Nucleare, Sezione di Ferrara, I-44100
Ferrara, Italy*

ABSTRACT

We extract information on the fluxes of Be and CNO neutrinos directly from solar neutrino experiments, with minimal assumptions about solar models. Next we compare these results with solar models, both standard and non standard ones. Finally we discuss the expectations for Borexino, both in the case of standard and non standard neutrinos.

## 1. Introduction

The principal aim of this paper is to extract information on the fluxes of Be and CNO neutrinos directly from solar neutrino experiments, with minimal assumptions about solar models. In this respect, we will update previous results[1-5] and try to elucidate the role of CNO neutrinos. We will see that experimental data are more and more against the hypothesis of standard neutrinos (*i.e.* without mass, mixing, magnetic moments...).

Next we will compare these informations with solar models, both standard and non standard ones. Clearly, low (*i.e.* smaller than standard) central temperature models are ruled out, essentially because they cannot reproduce the experimental data available on both Be and B neutrinos. Hybrid models, where some suitable nuclear cross section is varied in order to reduce the Be sneutrinos flux to the observed value and with a higher central temperature, so as to agree with experimental results on B neutrinos flux, can also be excluded, as in these models

---

the CNO neutrino flux grows beyond acceptable levels. In other words, the bounds on Be (CNO) neutrinos tell us that it is hopeless to reduce (enhance) the central solar temperature, in order to stay with standard neutrinos.

We will discuss then the expectations for Borexino, both in the case of standard and non standard neutrinos, showing that the experiment can clearly discriminate among several possible solutions to the solar neutrino puzzle.

## 2. Where are Be and CNO neutrinos ?

We make the assumption of stationary Sun (*i.e.* the presently observed luminosity equals the present nuclear energy production rate) and standard neutrinos, so that all the $\nu_e$ produced in the Sun reach Earth without being lost, and their energy spectrum is unchanged. The relevant variables are thus the (energy integrated) neutrino fluxes, which can be grouped as:

$$\Phi_{\text{pp+pep}}, \quad \Phi_{\text{Be}}, \quad \Phi_{\text{CNO}} \quad \text{and} \quad \Phi_{\text{B}} \ . \tag{1}$$

These four variables, see[2,3], are constrained by four equations:

a) the luminosity equation, which tells that the fusion of four protons (and two electrons) into one $\alpha$ particle is accompanied by the emission of two neutrinos, whichever is the cycle:

$$K_\odot = \sum_i \left(\frac{Q}{2} - \langle E \rangle_i\right) \Phi_i \tag{2}$$

where $K_\odot$ is the solar constant ($K_\odot = 8.5 \cdot 10^{11}$ MeV cm$^{-2}$ s$^{-1}$, Q=26.73 MeV) and $\langle E \rangle_i$ is the average energy of the i-th neutrinos.

b) The Gallium signal, $S_G$=(74 ± 8) SNU (weighted average between the Gallex[6] and SAGE[7] results), can be expressed as a linear combination of the $\Phi_i$'s, the weighting factors $\sigma_{i,G}$ being the absorption cross section for the i-th neutrinos, averaged on their energy spectrum, see[2]:

$$S_G = \sum_i \sigma_{i,G} \Phi_i \tag{3}$$

c) A similar equation holds for the Chlorine experiment, $S_C = (2.55 \pm 0.25)$SNU[8]:

$$S_C = \sum_i \sigma_{i,C} \Phi_i \tag{4}$$

d) The Kamiokande experiment determines - for standard neutrinos - the flux of Boron neutrinos[9]:

$$\Phi_B = (2.73 \pm 0.38) \cdot 10^6 \text{cm}^{-2}\text{s}^{-1}. \tag{5}$$

The four equations with four unknowns imply a unique solution:

$$\Phi_{Be} = (0.2 \pm 6.9) \cdot 10^9 \text{cm}^{-2}\text{s}^{-1}$$
$$\Phi_{CNO} = (-1.8 \pm 5.1) \cdot 10^9 \text{cm}^{-2}\text{s}^{-1}$$
$$\Phi_{pp+pep} = (66.4 \pm 8.2) \cdot 10^9 \text{cm}^{-2}\text{s}^{-1} \tag{6}$$

At first sight these results, taken at their face value and considering the errors, are not very clarifying about neutrino properties. However, let us observe that we have kept so far all the variables as free, without any restriction. On the other hand, if the hypotheses are correct the resulting fluxes are to be non negative, and this is not the case for the CNO-neutrinos.

In order to understand what is going on, and to make clear the role of each experimental result, let us reduce the number of equations and of unknowns by the following tricks:
a) one can eliminate $\Phi_{pp+pep}$ by using the luminosity equation (2).
b) Since $\langle E \rangle_{CNO} \geq \langle E \rangle_{Be}$, the corresponding cross section has to be larger than that of Be neutrinos. Thus the minimal CNO signal is obtained with the replacement

$$\sigma_{CNO} \to \sigma_{Be} \tag{7}$$

(We remark that this is also a safe approach, since the theoretical value of $\sigma_{Be,G}$ has essentially been verified to the 10% level by the Gallex neutrino source experiment[10]).

In this way, the above equations can be written in terms of two variables, $\Phi_{Be+CNO}$ and $\Phi_B$, and the results of each experiment can be plotted in the ($\Phi_B$, $\Phi_{Be+CNO}$) plane, see Fig. 1.

Clearly all experiments point towards $\Phi_{Be+CNO} < 0$. This means that the statement "neutrinos are standard and experiments are correct" has lead us to an unphysical conclusion. Could the problem be with some experiment? It is clear from Fig. 1 that the situation is unchanged by assuming that one experiment is wrong, see[11].

## 3. Experimental results and standard solar models

We have reported in Fig. 1 the results of several recent solar model calculations (diamonds)[12-19] together with experimental results. Some of the models predict a B flux close to the Kamiokande value; however no model is capable of reproducing the low Be+CNO flux implied by the experiments.

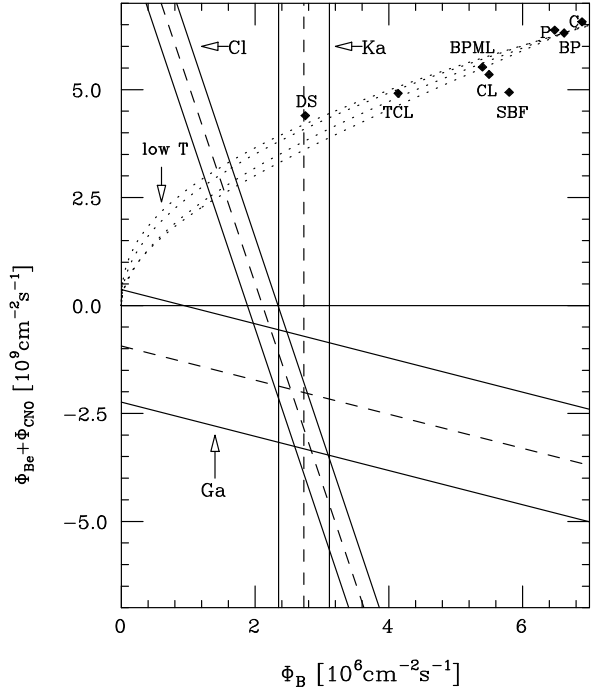

**Figure 1**

Neutrino fluxes allowed by the present experimental results. Dashed lines correspond to central values, solid lines denote $\pm 1\sigma$ limits. Diamonds represent recent solar model calculation[12-19]. The dotted area corresponds to (non-standard) low temperature solar models[2,20].

In Table 1, we have considered only standard solar models where He and heavier element diffusion is taken into account[13,15,19] as these should be more accurate. We remark that the comparison with helioseismology tells us that diffusion is important for solar models to predict the correct depth of the convective envelope[21,22]. We also note that in models with diffusion the central solar temperature is increased: as Helium falls towards the solar centre, the mean molecular weight increases in the stellar core and a higher temperature is needed to balance the gravitational force. Models with diffusion yield thus even larger Be, CNO and B neutrinos fluxes and the disagreement with respect to the experimental results is stronger.

Table 1. Information on Be, CNO and B neutrino fluxes. B neutrino flux in units of $10^6 cm^{-2} s^{-1}$; Be and CNO in units of $10^9 cm^{-2} s^{-1}$. Experimental bounds are at the 97% C. L.[28] . The remaining values are the results of solar models with diffusion: P94 from ref.[13] , BP95 from ref.[15] , FRANEC95 indicates our preliminary results[19] . The bound on CNO (last row) is obtained with "true" $\sigma_{CNO}$.

| flux | Exp. | P94 | BP95 | FRANEC95 |
|---|---|---|---|---|
| B | 2.73±0.4 | 6.48 | 6.62 | 6.9 |
| Be+CNO | ≤0.7 | 6.38 | 6.31 | 6.5 |
| Be | ≤0.7 | 5.18 | 5.15 | 5.3 |
| CNO | ≤0.6 | 1.20 | 1.16 | 1.2 |

The discrepancy between theory and experiment is about a factor two for the Boron flux. More important looks to us the discrepancy on $\Phi_{Be+CNO}$, where the predicted values are about an order of magnitude larger that the experimental upper bounds.

The discrepancy is essentially due to the Be neutrinos; note however that the upper bound on the sum is already below the SSM estimate for CNO. In other words, we have now a significant problem with intermediate energy (Be and CNO) neutrinos.

### 4. The relevance of Beryllium

As well known, theoretical predictions are more robust for Be- than for B- neutrinos, the reasons being the weaker sensitivity to the central solar temperature $T$ and the independence on the (poorly known) astrophysical factor $S_{17}$ for the $p+^7Be\rightarrow ^8B+\gamma$ reaction. Approximately, one has:

$$\Phi_{Be} = \Phi_{Be,0}(T/T_0)^{10}$$
$$\Phi_B = \Phi_{B,0}(S_{17}/S_{17,0})(T/T_0)^{20} \quad ,$$

(8)

where the subscript O refers here and in the following to the SSM predictions. For the power law coefficients see[2,23] .

In addition, we would like to point a relationship between $\Phi_{Be}$ and $\Phi_B$ which elucidateds physically the problem of the relative abundances of Be and B neutrinos.

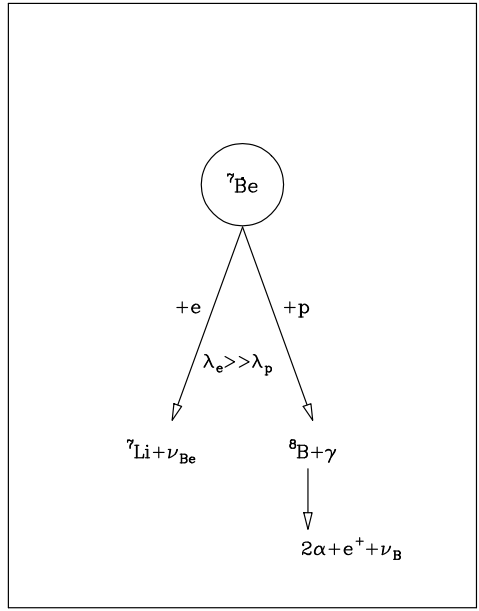

**Figure 2**

The fate of $^7$Be nuclei.

Both Be and B-neutrinos are sons of the $^7$Be nucleus, see Fig. 2. For this nucleus, electron capture (rate $\lambda_e$) is clearly favoured over proton capture (rate $\lambda_p$), due to the absence of the Coulomb barrier (it is curious that a weak process has a larger chance than an electromagnetic process, but this is the case due to the exponentially small penetration probabilities of the Coulomb barrier, at the energies of interest to us). Thus the value of $\Phi_{Be}$ is a clear indicator of the central density $n_7$ of the progenitors $^7$Be nuclei:

$$n_7 \propto \Phi_{Be}/\lambda_e \qquad (9)$$

If $\Phi_{Be}$ comes out to be smaller by an order of magnitude with respect to the SSM prediction, this implies the same for the $^7$Be equilibrium abundance (we recall that $\lambda_e$ is weakly dependent on temperature, and it is essentially known from measurements in the laboratory, see[24] ).
The puzzle is thus with B neutrinos, since:

$$\Phi_B \propto n_7 \lambda_p \qquad (10)$$

The observed (Kamiokande and Chlorine) value of $\Phi_B$ being just a factor two below the SSM prediction, the problem is now that experiments are observing too high $\Phi_B$! Put it in another way, one cannot kill the father/mother before the baby is conceived.

Should we insist on this road, we need to enhance $\lambda_p/\lambda_e$. We remark that any attempt to reduce $S_{17}$ goes into the wrong direction.

## 5. Reduced central temperature models?

Non standard solar models with smaller central temperaure can be obtained by varying - well beyond the estimated uncertainties - a few parameters (the cross section of the pp reaction, chemical composition, opacity, age...[2,20] ). These models span the dotted area in Fig. 1, which can be clearly understood by simple considerations.

To a rough approximation, also $\Phi_{CNO}$ has a power law dependence on temperature[2,23]:

$$\Phi_{CNO} = \Phi_{CNO,0}(T/T_0)^{20} \qquad (11)$$

One can use this equation together with eqs. (8) above; by expressing the temperature as a function of $\Phi_B$, one has:

$$\Phi_{Be} + \Phi_{CNO} = \Phi_{Be,0}(\Phi_B/\Phi_{B,0})^{1/2} + \Phi_{CNO,0}(\Phi_B/\Phi_{B,0}) \qquad (12)$$

and one sees in Fig. 1 the square root behaviour at small $\Phi_B$, which then changes to linear for larger $\Phi_B$.

It is clear that all these model fail to reproduce the experimental results, essentially because they cannot reproduce the observed ratio $\Phi_{Be}/\Phi_B$, see[25], as a consequence of the drastically different dependences on temperature, see eqs. 8. If $\Phi_B$ is reduced by a factor two, $\Phi_{Be}$ is too high. On the other hand, if $\Phi_B$ is brought to the low level required by the experiments, the predicted $\Phi_{Be}$ is definitely too small. In other words, as we said previously, we are observing too many B-neutrinos (if neutrinos are standard)!

## 6. Higher central temperatures? (Or why do we care about CNO neutrinos)

One could imagine the conspiracy of two mechanisms, so as to bring both $\Phi_{Be}$ and $\Phi_B$ in agreement with experiment. For example, one could assume that $S_{33}$ is much larger than commonly assumed (as a result of an hypothetical resonance[26]) so as to enhance the ppI channel and reduce $\Phi_{Be}$ to the desired value. At the same time, by varying some suitable parameter the central temperature could be increased, so as to bring $\Phi_B$ in agreement with experiment.

This mechanism also fails[27], see Fig. 3, due to the fact that as temperature raises, the CNO flux grows as fast as the Boron flux, and the experimental bound on $\Phi_{Be+CNO}$ is again violated.

In other words, while Beryllium and B neutrinos tell us that one cannot hope to solve the neutrino problem by lowering the central temperature, the bound on CNO implies that increasing the temperature does not work either.

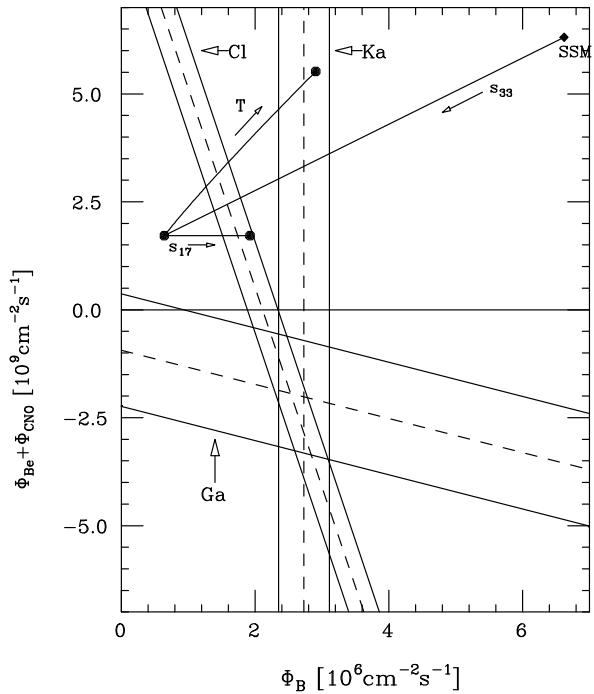

### Figure 3

Sketch of the behaviour of solar models with non standard $S_{33}$, central temperature T and $S_{17}$, from ref.[27].

If instead the temperature is unchanged but $S_{17}$ is increased, still one has the problem that the SSM prediction for the CNO exceeds the experimental bound, see again Fig. 3 and ref.[27].

## 7. Expectations for Be neutrinos

We have seen that, for standard neutrinos, the Be-flux is strongly suppressed with respect to the SSM predictions. What has to be expected for non standard neutrinos?

In Table 2 we show the results of a recent analysis[28], for a few candidate solutions. It is worth observing that, at least for active neutrinos, both small and large angle MSW solutions are acceptable, as well as the Just-So model. From the last colum, we see that the signal (CC+NC) corresponding to MSW large angle and to the Just-So models best fit points are quite a significant fraction of the standard model prediction. In other words, in face of the present experimental data, the Beryllium signal does not need to be small (for non standard neutrinos).

Table 2. Predictions for Beryllium neutrinos. For different models we present the best fit point ($\sin^2\theta$, $\delta m^2$), the $\chi^2$ for degree of freedom, the flux and signal (CC+NC) in units of the SSM predictions (from ref.[28]).

|  | $\chi^2/d.o.f.$ | $\sin^2\theta$ | $\delta m^2$ | $\Phi/\Phi_0$ | $S/S_0$ |
|---|---|---|---|---|---|
| MSW small $\theta$ | 0.5/2 | .0058 | 6.24 | 3% | 22% |
| MSW large$\theta$ | 2.7/2 | 0.73 | 29.9 | 54% | 64% |
| just-so | 2.4/2 | 0.86 | 6.15 | 48% | 59% |

The situation is made more clear in Fig. 4, where we show the 90% CL regions according to the different models. It is interesting to observe that a direct measurement of the Be-line can in many cases discriminate among the possibile solutions. Very large signals, above 75% of the SSM prediction, correspond essentially to the Just-So solution. Between 75% and about 35% various models are acceptable. Between 35% and 20% the solution has to bee MSW at small angle. Very small signal, say below 20%, are only possible for standard neutrinos (or transitions into sterile neutrinos).

In the intermediate region discrimination between Just-So and MSW solution can be obtained by Borexino looking at seasonal variations, even for purities well below the design purity, see[28].

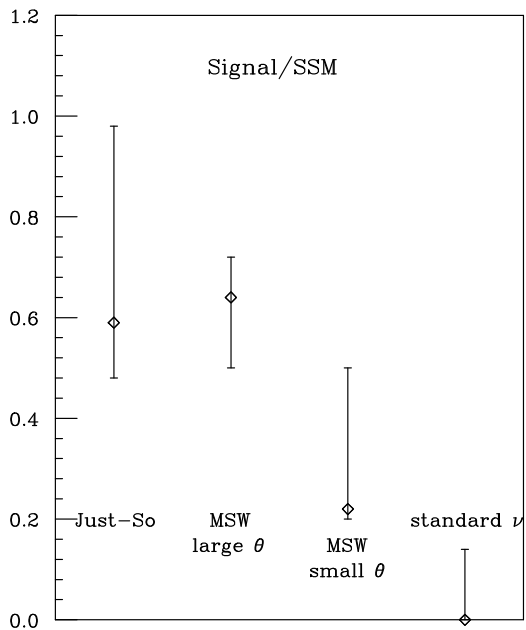

**Figure 4**

The Beryllium (CC+NC) signal, in units of the SSM prediction. Diamonds indicate the best fit points, bars correspond to 90 % C.L. (from ref.[28]).

***

Most of the results presented here come from the fruitful and friendly collaboration in the last few years with Berezinsky, Calabresu, Castellani, Ciacio, Degl'Innocenti, Ferrari and Lissia.